\documentclass[aps,pra,twocolumn,a4paper,showpacs,superscriptaddress,10pt]{revtex4-1}

\usepackage{amsmath,amssymb}
\usepackage{ulem,graphicx,xcolor}
\usepackage[colorlinks=true,urlcolor=blue,citecolor=blue,linkcolor=blue]{hyperref}

\begin{document}
\title{Gaussian entanglement in the turbulent atmosphere}

\author{M. Bohmann}\email{martin.bohmann@uni-rostock.de}
\affiliation{Institut f\"ur Physik, Universit\"at Rostock, Albert-Einstein-Str. 23, D-18051 Rostock, Germany}

\author{A. A. Semenov}
\affiliation{Institut f\"ur Physik, Universit\"at Rostock, Albert-Einstein-Str. 23, D-18051 Rostock, Germany}
\affiliation{Institute of  Physics, NAS of Ukraine, Prospect Nauky 46, UA-03028 Kiev, Ukraine}

\author{J. Sperling}
\affiliation{Institut f\"ur Physik, Universit\"at Rostock, Albert-Einstein-Str. 23, D-18051 Rostock, Germany}
\affiliation{Clarendon Laboratory, University of Oxford, Parks Road, Oxford OX1 3PU, United Kingdom}

\author{W. Vogel}
\affiliation{Institut f\"ur Physik, Universit\"at Rostock, Albert-Einstein-Str. 23, D-18051 Rostock, Germany}

\begin{abstract}
	We provide a rigorous treatment of the entanglement properties of two-mode Gaussian states in atmospheric channels by deriving and analyzing the input-output relations for the corresponding entanglement test.
	A key feature of such turbulent channels is a non-trivial dependence of the transmitted continuous-variable entanglement on coherent displacements of the quantum state of the input field.
	Remarkably, this allows one to optimize the entanglement certification by modifying local coherent amplitudes using a finite, but optimal amount of squeezing.
	In addition, we propose a protocol which, in principle, renders it possible to transfer the Gaussian entanglement through any turbulent channel over arbitrary distances.
	Therefore, our approach provides the theoretical foundation for advanced applications of Gaussian entanglement in free-space quantum communication.
\end{abstract}

\date{\today}

\maketitle

\section{Introduction}
	Based on the fundamental principles of quantum mechanics, quantum protocols can increase the security of communication channels \cite{Gisin}.
	Quantum-based communication systems using optical fibers are already commercially available.
	However, one faces the disadvantages of limited flexibility concerning the positions of sender and receiver, bounding the distances to $\sim$100~km, due to losses \cite{Fibre1,Fibre2,Fibre3,Fibre4}.
	An alternative consists in atmospheric free-space channels.
	In past years, it has been demonstrated that quantum communication is possible through free-space links \cite{Ursin,Fedrizzi} and even via orbiting satellites \cite{Satellite0,Satellite2,Satellite3,Satellite4,Satellite5}.
	This opens the possibility to establish a global quantum communication network.

	Beside many experimental demonstrations in discrete variables \cite{Ursin,Fedrizzi,Satellite0,Satellite2,Satellite3}, a different approach is the continuous-variable quantum key distribution (CV-QKD) \cite{CV-QKD1,CV-QKD2,CV-QKD3,CV-QKD4,CV-QKD5,CV-QKD6}.
	The latter works even in the presence of bright day light \cite{Elser,Heim,Peuntinger}.
	However, the standard CV-QKD protocols require further improvements, as current methods only support rather limited communication distances \cite{Fossier,Xuan,Jouguet2012,Jouguet2013}.

	The generation of Gaussian entangled states is nowadays quite advanced and, thus, they serve as the main class of states for improving CV-QKD \cite{Rodo,DiGuglielmo,Su,Madsen,Furrer,Eberle2013}.
	Gaussian states are fully characterized by the covariance matrix of their field quadratures or, equivalently, of their complex field amplitudes. 
	Bipartite Gaussian entanglement can be always uncovered by the Simon criterion \cite{Simon2000}, which is based on the partially transposed covariance matrix. 
	Another criterion by Duan {\it et al.} \cite{Duan2000} is necessary and sufficient for a particular choice of the computational or measurement basis. 
	
	The description of the quantum properties of light after transmission through the turbulent atmosphere requires a quantum theory of atmospheric fading channels with fluctuating losses \cite{Semenov2009,Semenov2010,beamwandering,VSV2016}.
	Such channels are characterized by a probability distribution of transmission (PDT).
	The derivation of the PDT relies on detailed knowledge of the atmospheric properties, such as the propagation distance, the weather, and the daytime conditions. 
	It requires one to unify the knowledge on classical atmospheric optics~\cite{Tatarskii} with quantum optics~\cite{VSV2016}.

	For Gaussian quantum light, one usually measures the field quadratures with balanced homodyne detection.
	This technique has been adapted for fading atmospheric channels \cite{Elser,Heim,Peuntinger,Semenov2012} by propagating the signal and the reference field, i.e., the local oscillator, in orthogonal polarization modes.
	Some scenarios with Gaussian entanglement in the atmosphere have been studied \cite{Usenko,GaussSatellites}, but a full analysis of the evolution of bipartite Gaussian entanglement in arbitrary free-space links is missing yet.
	Such a complete analysis would allow one to optimize the transmitted entanglement, e.g., for ensuring a maximal security of CV-QKD protocols.
	The treatment of secure data transfer through atmospheric channels needs interdisciplinary research, combining the fields of classical atmospheric optics, quantum optics, and quantum information theory.

	In this contribution, we aim at a full study of the transmission of bipartite Gaussian entanglement through the turbulent atmosphere.
	For this purpose, we introduce input-output relations for the Simon entanglement criterion.
	We consider two fundamentally different cases. 
	The first one is the case of uncorrelated fading channels where both modes are subjected to independent losses, e.g., due to propagation in different directions.
	The second one is the case of correlated fading channels.
	Based on adaptive methods, we show that any channel can attain this correlated form, which results in entanglement preserving links.
	A remarkable consequence of our analysis is a dependence of the Gaussian entanglement on the local coherent displacements of the input fields.
	An adjustment of these parameters allows one to optimize Gaussian entanglement transfer through the turbulent atmosphere. 
	
	The article is structured as follows.
	In Sec.~\ref{ch:general}, we derive the output covariance matrix and the corresponding entanglement test for the transmission of bipartite Gaussian entanglement trough turbulent channels.
	We focus on the effects of uncorrelated atmospheric losses in Sec.~\ref{ch:uncorrelated}.
	In Sec.~\ref{ch:adaptive}, we introduce an adaptive method to correlate the channels and discuss the Gaussian entanglement in this case.
	A summary and conclusions are given in Sec.~\ref{ch:summary}.

\section{Gaussian entanglement in fading channels}\label{ch:general}
	Gaussian states are completely described by the first- and second-order moments of their field quadratures or, equivalently, bosonic creation and annihilation operators.
	The Simon entanglement criterion \cite{Simon2000} in the form of Ref. \cite{ShchukinVogel2005} states that any two-mode Gaussian state is entangled if and only if
	\begin{equation} \label{Witness}
		\mathcal{W}=\det V^{\mathrm{PT}}<0,
	\end{equation}
	where $V^{\mathrm{PT}}$ is the partial transposition of the matrix
	\begin{align}\label{eq:SimonMatrix}
	\begin{aligned}
		V{=}&
		\begin{pmatrix}
			\langle \Delta\hat a^\dagger\Delta \hat a\rangle & \langle \Delta\hat a^{\dagger2}\rangle & \langle \Delta\hat a^\dagger\Delta \hat b\rangle & \langle \Delta\hat a^\dagger\Delta \hat b^\dagger\rangle \\
			\langle \Delta\hat a^2\rangle & \langle \Delta\hat a\Delta\hat a^\dagger\rangle & \langle \Delta\hat a\Delta \hat b\rangle & \langle \Delta\hat a\Delta \hat b^\dagger\rangle \\
			\langle \Delta\hat a\Delta\hat b^\dagger\rangle & \langle \Delta\hat a^\dagger\Delta\hat b^\dagger\rangle & \langle \Delta\hat b^\dagger\Delta \hat b\rangle & \langle \Delta \hat b^{\dagger2}\rangle \\
			\langle \Delta\hat a\Delta\hat b\rangle & \langle \Delta\hat a^\dagger\Delta\hat b\rangle & \langle \Delta\hat b^2\rangle & \langle \Delta \hat b\Delta \hat b^\dagger\rangle
		\end{pmatrix}
		\\{=}&\begin{pmatrix}A&C^\dagger\\C&B\end{pmatrix}.
	\end{aligned}
	\end{align}
	Here, $V$ is the second-order matrix in bosonic creation and annihilation operators, with $\hat{a}$ and $\hat{b}$ denoting the annihilation operators of the two field modes and $\Delta\hat x=\hat x-\langle\hat x\rangle$, with $\hat x=\hat a, \hat b$.
	The matrix $V$ can be given in a block form with $2\times2$ blocks $A$, $B$, and $C$, where $A$ and $B$ are related to the single-mode covariances and $C$ describes the correlations between the modes.
	The description with bosonic field-mode operators can always be rewritten in terms of quadratures via a linear transformation.
	Note that the Simon criterion~Eq.~\eqref{Witness} is an entanglement test which indicates entanglement. 
	However, it is not aimed at quantifying the amount of entanglement.
	For a quantification of Gaussian entanglement, one needs to employ entanglement measures (monotones); see, e.g.,~\cite{SSV13}.

	In the following, we will introduce the treatment of atmospheric fading channels.
	The theoretical description of a general two-mode quantum state after transmission through fading channels is a bipartite generalization of the theory established in Ref. \cite{Semenov2009}.
        In this context, the elements of matrix~(\ref{eq:SimonMatrix}) can be expressed in terms of normally ordered moments,
        \begin{equation}\label{MomentForm}
		\langle \hat{a}^{\dagger n}\hat{a}^{m}\hat{b}^{\dagger k}\hat{b}^{l}\rangle_\mathrm{atm.}
		=\langle T_a^{n+m}T_b^{k+l}\rangle\langle \hat{a}^{\dagger n}\hat{a}^{m}\hat{b}^{\dagger k}\hat{b}^{l}\rangle.
        \end{equation}
        Here, $\langle \hat{a}^{\dagger n}\hat{a}^{m}\hat{b}^{\dagger k}\hat{b}^{l}\rangle$ are the field moments of the state and $\langle \hat{a}^{\dagger n}\hat{a}^{m}\hat{b}^{\dagger k}\hat{b}^{l}\rangle_\mathrm{atm.}$ denote the moments after propagating the state through the turbulent atmosphere.
	Moreover, $T_a$ and $T_b$ denote the amplitude transmission coefficients of the two field modes which are fluctuating according to the joint PDT, $\mathcal{P}(T_a,T_b)$, i.e., 
	\begin{align}
		\langle T_a^{n+m}T_b^{k+l}\rangle=\int_0^1 dT_a\int_0^1 dT_b\,\mathcal P(T_a,T_b) T_a^{n+m}T_b^{k+l}.
	\end{align}
	Due to the negligibly small depolarization effects of the atmosphere \cite{Tatarskii} and no dephasing between different polarizations \cite{Elser,Heim,Peuntinger}, the transmission coefficients can be considered as real random variables \cite{Semenov2012}.
	In contrast to fading channels with fluctuating loss, deterministic loss channels are characterized by a deterministic PDT, $\mathcal P(T_a,T_b)=\delta(T_a-\sqrt{\eta_a})\delta(T_b-\sqrt{\eta_b})$, where $1-\eta_{a(b)}$ denotes the constant loss in the subsystem $a(b)$. For example,
	deterministic losses may properly describe the attenuation in optical fibers.
	
	The above description of fading losses is general and applies to all passive polarization-preserving turbulent loss media.
	However, in order to give some realistic examples in the further course of this work, we will shortly comment on existing models of different atmospheric loss regimes which show good agreement with experimental data.
	In the case of weak turbulence, the leading effect is beam wandering, which describes the wandering of the beam spot at the receiver aperture plane due to atmospheric turbulence.
	For this effect, a PDT model has been derived \cite{beamwandering}, which agrees well with experiments with a propagation distance of  $1.6$ km~\cite{Usenko}.
	Recently, the approach was generalized for including elliptic beam-shape deformation effects \cite{VSV2016}.
	This model even applies to conditions of strong turbulence. The derived PDT behaves similar to that in Ref. \cite{Villoresi2012} for a $144$ km free-space link. 
	These examples show that the propagation distance is one of the relevant parameters controlling the transition from weak to strong turbulence.

	After passing through turbulent channels, the quantum state of light is, in general, not Gaussian anymore. 
	The Simon criterion is solely based on moments of second order.
	Hence, it yields a complete entanglement characterization of bipartite Gaussian states. 
	For non-Gaussian states, it still certifies the entanglement inherent in the second moments, denoted as the Gaussian part of entanglement.
	However, it cannot identify entanglement effects related to higher-order correlations.
	Let us formulate the input-output relations, which connect matrix~\eqref{eq:SimonMatrix} at the source with that at the receivers.
	For technical details, we refer to~\cite{Supplementary}.
	The atmospheric output matrix reads
	\begin{align}\label{IOR_V}
		V_{\rm atm.}=V_{\langle T_{a}^2\rangle,\langle T_{b}^2\rangle,\Gamma}
		+\begin{pmatrix}
			\vec\mu_a\vec\mu_a^\dagger &
			\Delta\Gamma \vec\mu_a\vec\mu_b^\dagger \\
			\Delta\Gamma \vec\mu_b\vec\mu_a^\dagger &
			\vec\mu_b\vec\mu_b^\dagger
		\end{pmatrix}.
	\end{align}
	In the following, we discuss this result in detail including the notations and the used symbols.

	The effect of a deterministic loss, $1-\eta_{a(b)}$, is represented by the attenuated matrix $V_{\eta_a,\eta_b,1}$, which has been extensively studied \cite{Filippov}.
	Here, we have $\eta_{a(b)}=\langle T_{a(b)}^2\rangle$.
	The atmospheric matrix $V_{\rm atm.}$ is further characterized by the two correlation coefficients
	\begin{align}\label{CorrPara}
			\Gamma=\frac{\langle T_a T_b\rangle}{\sqrt{\langle T_a^2\rangle\langle T_b^2\rangle}}
			\quad\text{and}\quad
			\Delta\Gamma=\frac{\langle \Delta T_a\Delta T_b\rangle}{\sqrt{\langle \Delta T_a^2\rangle\langle \Delta T_b^2\rangle}}.
	\end{align}
	It is easy to see that $\Gamma,|\Delta\Gamma|\in[0,1]$ \cite{CSI}.
	The $\Gamma$ index in the matrix $V_{\eta_a,\eta_b,\Gamma}$ [cf. Eq.~\eqref{IOR_V}] indicates that the correlations between the two modes are diminished by a factor $\Gamma<1$.
	That is, the correlation block in Eq.~\eqref{eq:SimonMatrix} maps as $C\mapsto\Gamma C$.
	Moreover, the second term in Eq.~\eqref{IOR_V} depends on the vectors of local displacement $\vec\mu_a=\sqrt{\langle \Delta T_a^2\rangle}(\langle \hat a^\dagger\rangle,\langle\hat a\rangle)^{\rm T}$ and $\vec\mu_b=\sqrt{\langle \Delta T_b^2\rangle}(\langle \hat b^\dagger\rangle, \langle\hat b\rangle)^{\rm T}$.
	Note that these vectors are scaled with the atmospheric fluctuations of the transmission coefficients.
	Hence, fading channels lead to a strict dependence of the entanglement certification on the coherent amplitudes, unlike in the case of deterministic attenuation ($\langle \Delta T_{a}^2\rangle=\langle \Delta T_{b}^2\rangle=0$).
	Again, the off-diagonal (intermode) part of the displacement contribution in Eq.~\eqref{IOR_V} is scaled with a correlation coefficient, $\Delta\Gamma$.
	Thus, major channel characteristics are determined by the two correlation parameters $\Gamma$ and $\Delta\Gamma$, which will be studied for different scenarios later on.

	The corresponding entanglement criterion, $\mathcal{W}_{\rm atm.}=\det V_{\rm atm.}^{\rm PT}<0$, is calculated by inserting the partial transposition of $V_{\rm atm.}$ in  Eq.~\eqref{IOR_V} into Eq.~\eqref{Witness}.
	This results in~\cite{Supplementary}
	\begin{align}\label{eq:WitnessIOR}
	\begin{aligned}
		\mathcal{W}_{\rm atm.}=&\Gamma^2\mathcal{W}_{\langle T_{a}^2\rangle,\langle 
		T_{b}^2\rangle,1}+(1-\Gamma^2)\mathcal{N}
		\\&+(1-\Delta\Gamma^2)\mathcal{F}+
		\vec\nu^\dagger \mathcal{S}\vec \nu,
	\end{aligned}
	\end{align}
	where $\mathcal{W}_{\langle T_{a}^2\rangle,\langle T_{b}^2\rangle,1}=\det V_{\langle 
	T_{a}^2\rangle,\langle T_{b}^2\rangle,1}^{\rm PT}$ is the well-known deterministic loss contribution
	and with a displacement vector $\vec \nu=(\vec\mu_{b\perp}^\dagger, -\vec\mu_{a\perp}^T)^T$ ($\vec x_\perp$ indicates the perpendicular vector to $\vec x$).
	Additionally, we have the terms
	\begin{align}\nonumber
		\mathcal{N}{=}&\det\begin{pmatrix}
			\det \tilde A & \Gamma\det\tilde C^\dagger\\\Gamma\det\tilde C&\det\tilde B
		\end{pmatrix}
		\!,\,
		\mathcal{S}{=}\begin{pmatrix}
			S_{aa} & \Gamma \Delta\Gamma S_{ba}^\dagger \\
			\Gamma\Delta\Gamma S_{ba} & S_{bb}
		\end{pmatrix}\!,\,
		\\&\text{and }\label{Eq:NSF}
		\mathcal{F}{=}\det\begin{pmatrix}
			\vec{\mu}^\dagger_{a\perp}\tilde A\vec\mu_{a\perp} & \Gamma\vec\mu^\dagger_{a\perp}\tilde C^\dagger\vec\mu_{b\perp}^\ast \\
			\Gamma\vec\mu^{\rm T}_{b\perp}\tilde C\vec\mu_{a\perp} & \vec\mu^{\rm T}_{b\perp}\tilde B\vec\mu_{b\perp}^\ast
		\end{pmatrix},
	\end{align}
	where $\tilde X$ (for $X{=}A,B,C$) denotes the $2\times 2$ blocks of $V_{\langle T_{a}^2\rangle,\langle T_{a}^2\rangle,1}$ [cf. Eqs.~\eqref{eq:SimonMatrix} and~\eqref{IOR_V}] after partial transposition, and
	$S_{aa}=\det\tilde A(\tilde B-\Gamma^2\tilde C\tilde A^{-1}\tilde C^\dagger)$,
	$S_{ba}=-\det\tilde C(\Gamma^2\tilde C^\dagger-\tilde A\tilde C^{-1}\tilde B)$,
	as well as $S_{bb}=\det \tilde B(\tilde A-\Gamma^2\tilde C^\dagger \tilde B^{-1}\tilde C)$.

	Let us analyze the structure of Eq.~\eqref{eq:WitnessIOR}.
	The first two terms represent the decrease of the correlation between the two modes by the factor $\Gamma<1$ in turbulent loss channels [see Eqs.~\eqref{IOR_V} and~\eqref{CorrPara}] where $\mathcal{N}>0$.
	The last two terms in Eq.~\eqref{eq:WitnessIOR}, being related to $\mathcal{F}\geq 0$ and $\mathcal{S}$, show the dependency of the Simon entanglement test on coherent displacements, which is an important finding for turbulent loss channels.
	In particular, the contribution including $\vec \nu^\dagger\mathcal{S}\vec \nu$ can be negative for $\Delta\Gamma\neq 0$ with proper choices of displacement vectors.
	This leads to the fact that the entanglement transfer can be optimized, as we will discuss in the continuation of this work.
	Note that, in the case of deterministic attenuation, i.e., $\langle\Delta T_{a(b)}^2\rangle=0$, this criterion reduces to the case of deterministic attenuation which always preserves Gaussian entanglement \cite{Filippov,ConsatntLoss1}.
	Equations~\eqref{IOR_V} and~\eqref{eq:WitnessIOR} represent the most general form of the input-output relation for the Gaussian entanglement test in atmospheric links.

\section{Uncorrelated fading channels} \label{ch:uncorrelated} 

	After this full treatment, we continue our analysis with the case of uncorrelated channels, with $\langle T_a^mT_b^n\rangle=\langle T_a^m\rangle\langle T_b^n\rangle$.
	Thus, we have for the correlation parameters in Eq.~\eqref{CorrPara}: $\Gamma=\big(\langle T_a\rangle/\sqrt{\langle T_a^2\rangle}\big)\big(\langle T_b\rangle/\sqrt{\langle T_b^2\rangle}\big)$ and $\Delta\Gamma=0$.
	A natural example is the case of counterpropagation, i.e., both modes propagate in different directions through the atmosphere.
	The case in which one mode undergoes only a deterministic loss is also included.
	An example is a scenario where entanglement is established between the sender, who locally keeps one mode in a fiber loop, and a remote receiver of the other mode, connected via a free-space link \cite{Ursin}.

	Let us assume zero coherent displacements, $0=\vec\mu_a=\vec\mu_b=\vec\nu$.
	Then, the entanglement test~\eqref{eq:WitnessIOR} reduces to 
	\begin{align}\label{eq:unnoshift}
		\mathcal{W}_{\rm atm.}=\Gamma^2\mathcal{W}_{\langle T_{a}^2\rangle,\langle T_{b}^2\rangle,1}+(1-\Gamma^2)\mathcal{N}.
	\end{align}
	The first term resembles the deterministic loss scaled by the atmospheric coefficient $\Gamma^2$ [cf. Eq.~\eqref{CorrPara}]. 
	The second term is clearly positive, as $\mathcal{N}>0$.
	With decreasing $\Gamma$, the absolute value of the negative first term becomes smaller, while the positive second term increases. 
	Consequently, the Gaussian entanglement part vanishes, as $\mathcal{W}_{\rm atm.}$ becomes positive.

\begin{figure}[h!]
	\includegraphics[width=0.9\linewidth]{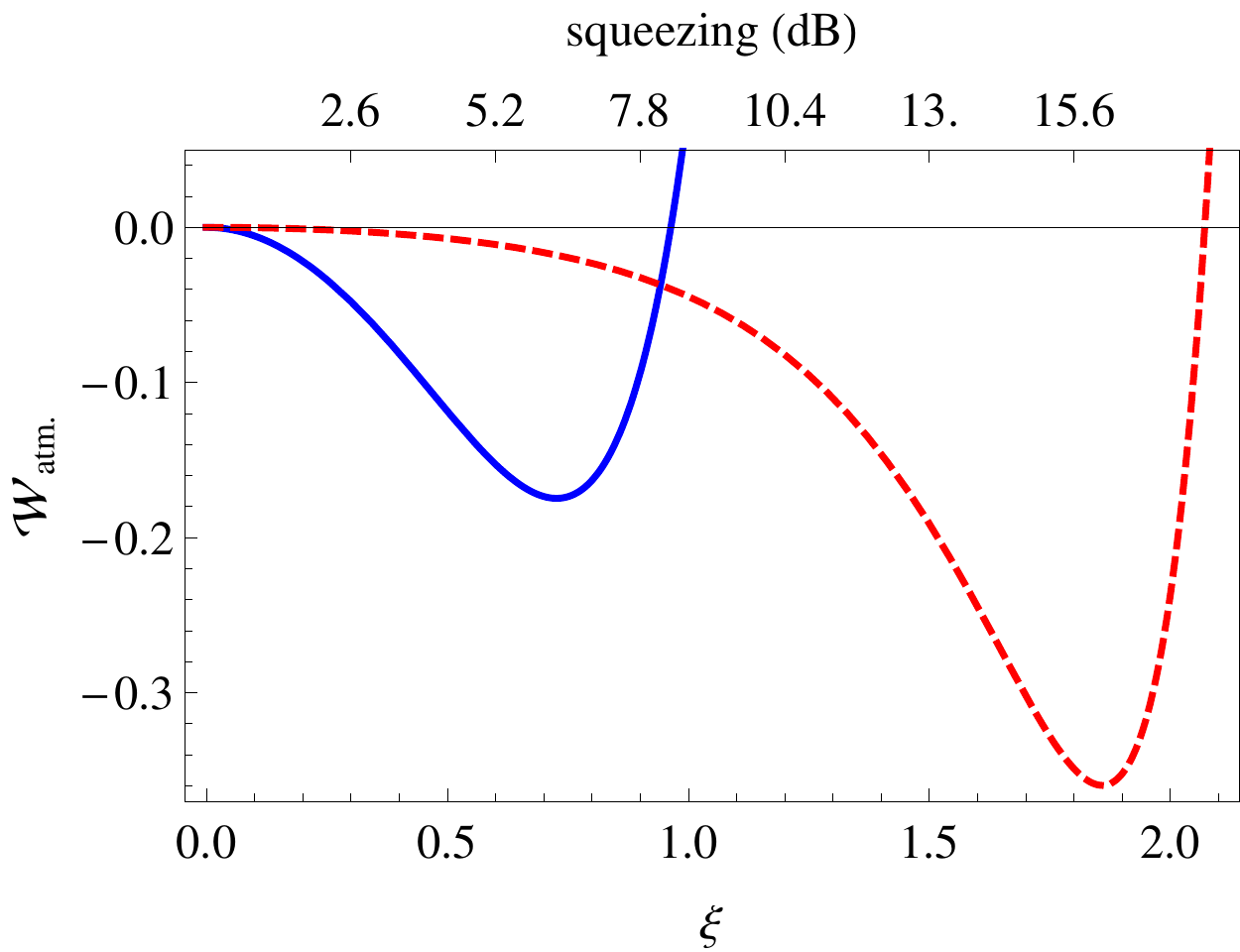}
	\caption{(Color online)
		The Simon entanglement test $\mathcal{W}_\mathrm{atm.}$ is shown for a TMSV state for two uncorrelated atmospheric channels with channel characteristics obtained from experiments.
		The solid and dashed lines correspond to a 144-km-long channel with 
		$\langle T_a\rangle{=}\langle T_b\rangle{=}0.027$, $\langle T_a^2\rangle{=}\langle 
		T_b^2\rangle{=}0.001$ ($\Gamma{\approx}0.75$) and to a 1.6-km-long channel with 
		$\langle T_a\rangle{=}\langle T_b\rangle{=}0.398$, $\langle T_a^2\rangle{=}\langle 
		T_b^2\rangle{=}0.163$ ($\Gamma{\approx}0.97$), respectively.
		$\mathcal{W}_\mathrm{atm.}$ is scaled by $10^6$ for the solid line.
		Surprisingly, if $\xi$ exceeds a certain value no Gaussian entanglement persists.
	}\label{fig:tmsv}
\end{figure}
	
	A surprising effect can be observed when we consider the propagation of a two-mode squeezed-vacuum (TMSV) state through uncorrelated atmospheric channels;
	$|{\rm TMSV}\rangle=(\cosh\xi)^{-1}\sum_{n=0}^{\infty}(\tanh \xi)^n |n,n\rangle$ with squeezing parameter $\xi\geq0$.
	In Fig.~\ref{fig:tmsv}, we see that the increase of squeezing can frustrate the transfer of Gaussian entanglement through atmospheric links.
	Thus, too strong squeezing might be hindering in turbulent loss channels.
	Consequently, an optimal squeezing interval can be identified.
	Let us stress that this statement has to be understood in terms of the significance of verified entanglement, which yields an operational quantification of the entanglement transferred through the atmosphere in the sense of its significance.
	We examine the initial partially transposed matrix $V^{\rm PT}$ of the TMSV state.
	All blocks of $V^{\rm PT}$ have a diagonal form, $\tilde A=\tilde B={\rm diag}(\sinh^2\xi,\cosh^2\xi)$ and $\tilde C={\rm diag}(\sinh \xi \cosh \xi,\sinh \xi \cosh \xi)$. 
	With increasing $\xi$, all nonzero entries of $V^{\rm PT}$ become approximately $e^{2\xi}$.
	However, the turbulence reduces the correlations $\tilde C$ by the factor $\Gamma$.
	As these correlations are responsible for the entanglement this reduction eventually yields $\mathcal{W}_{\rm atm.}>0$.
	A more intuitive way to understand this effect can be given by considering the squeezing ellipses.
	A TMSV state appears to show squeezing in the joint position and momentum basis of the two modes, respectively.
	The stronger the squeezing the more pronounced are these ellipses.
	Uncorrelated turbulent losses cause different fluctuations in both modes which leads to rotational blurring of the original squeezing ellipses around the origins.
	For stronger squeezing this smearing effect is more grave as the antisqueezed parts of the ellipses fluctuate more intensively.
	This leads to the fact that TMSV states with stronger squeezing may disentangle faster in uncorrelated turbulent loss channels.
	In particular, the smaller $\Gamma$ is, the smaller is the squeezing interval for which Gaussian entanglement survives (cf.~Fig.~\ref{fig:tmsv}).
	
	In order to exemplarily demonstrate the capability of our approach to describe real atmospheric channels, we use measured experimental transmission characteristics.
	In Fig.~\ref{fig:tmsv} we apply turbulent loss parameters obtained from two experiments which correspond to a rather long and a short transmission channel:
	the first one for a 144-km-long free-space link between two Canary islands \cite{Villoresi2012} and the second one for a 1.6-km-long atmospheric channel \cite{Usenko}.
	For the former, we used a PDT model which resembles the log-normal distribution \cite{VSV2016}, the latter is described by the dominant effect of beam wandering \cite{beamwandering}.
	Both PDT models are in good agreement with the corresponding experimental data \cite{Villoresi2012,Usenko}, which demonstrates the applicability for different atmospheric conditions.

	Now, we will consider the surprising effect of coherent displacements on the Simon test for uncorrelated fading channels. 
	As $\Delta\Gamma=0$, we see that the second term in Eq.~\eqref{IOR_V} only adds a positive part to the single-mode blocks of the matrix $V_{\rm atm.}$, i.e., $A$ and $B$ get the additional summands $\vec\mu_a\vec\mu_a^\dagger$ and $\vec\mu_b\vec\mu_b^\dagger$, respectively, and $C$ has no additional displacement dependent term.
	In terms of the test in Eq.~\eqref{eq:WitnessIOR}, this means that the last term is positive and increases with the local displacements.
	As a consequence, Gaussian entanglement vanishes with increasing the values of $\langle \hat{a}\rangle$ and $\langle \hat{b}\rangle$.
	Here, a similar argument, as given above for the frustration of entanglement transfer by strong squeezing, can be employed to give the reader a more intuitive explanation of this behavior. 
	We already mentioned that uncorrelated turbulent losses lead to rotational blurring in the phase space which is more pronounced the further the state is displaced from the origin.
	Consequently, one can directly understand why coherent displacements are disadvantageous in such environments. 
	Hence, coherent displacements should be avoided to preserve entanglement in uncorrelated atmospheric channels.
	However, for certain scenarios, such as CV-QKD protocols, one employs coherent displacement (see, e.g., \cite{Madsen,Ralph}).
	In such cases, one may optimize the displacements to conserve the entanglement.

\begin{figure}[h!]
	\includegraphics[width=0.8\linewidth]{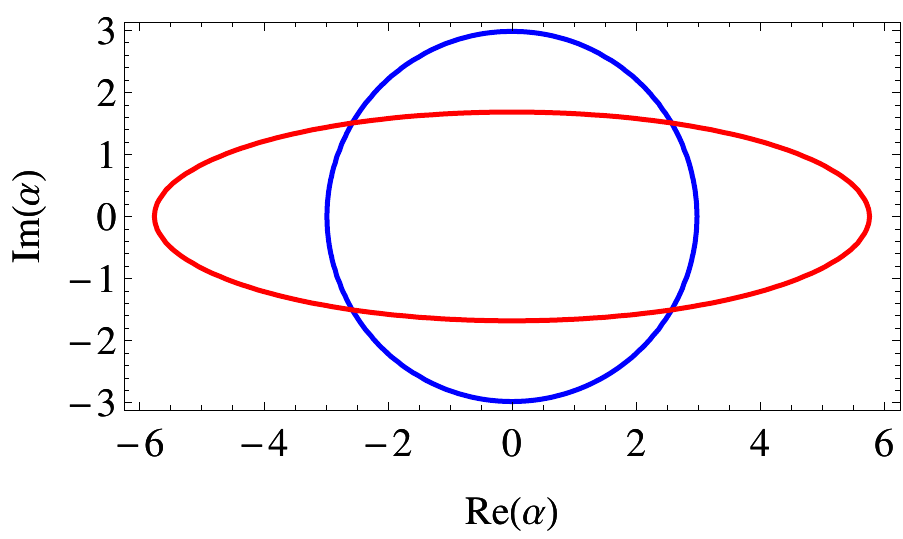}
	\caption{(Color online)
		The roots of the Simon entanglement test, $\mathcal W_{\rm atm.}{=}0$, are displayed as a function of the coherent displacement $\alpha$.
		The circle and the ellipse correspond to the displaced symmetric and asymmetric TMSV states with squeezing parameter $\xi=1$.
		The interior (exterior) regions satisfy $\mathcal W_{\rm atm.}{<}0$ ($\mathcal W_{\rm atm.}{>}0$).
		The channel parameters are  $\langle T_a\rangle{=}\langle T_b\rangle{=}\langle T_{a}^2\rangle{=}\langle T_b^2\rangle{=}0.9$ ($\Gamma{=}0.9$).
	}\label{fig:anisotropy}
\end{figure}

	An example of the displacement dependence in phase space is shown in Fig.~\ref{fig:anisotropy}.
	We study an asymmetric TMSV state, which can be generated by mixing two equally squeezed, single-mode states on a beam splitter with transmission coefficient of $t^2=0.95$ and a consecutive displacement in one mode, $\alpha=\langle\hat a\rangle\neq0$ and $\langle\hat b\rangle=0$.
	It is worth mentioning that the standard TMSV state is obtained with a $50{:}50$ beam splitter ($t^2=0.5$).
	The here considered asymmetry of the input state leads to larger values of $\alpha$ in some directions of the phase space for which the entanglement is still preserved.

\section{Adaptive channel correlations}\label{ch:adaptive}

	We proceed with our analysis to characterize the case of fully correlated channels, i.e., $\langle T_a^mT_b^n\rangle=\langle T_a^{m+n}\rangle=\langle T_b^{m+n}\rangle$.
	Thus, we get the maximal values $\Gamma=\Delta \Gamma=1$ [see Eq.~\eqref{CorrPara}] which reduces Eq.~\eqref{eq:WitnessIOR} to
	\begin{align}\label{PerfectCorrWitness}
		\mathcal{W}_{\rm atm.}=\mathcal{W}_{\langle T_{a}^2\rangle,\langle T_{a}^2\rangle,1}+\vec\nu^\dagger \mathcal{S}\vec\nu. 
	\end{align}
	Completely correlated fading channels can be established in the case of copropagation \cite{Fedrizzi, Semenov2010}. 

	Alternatively, this ideal correlation can be produced in other kinds of communication channels by artificially monitoring and adapting the channel transmissivities.
	In detail, (i) one has to measure the transmission coefficients in both channels, (ii) share this information via classical communication, and (iii) attenuate the channel with the higher transmissivity to the level of the lower one.
	The online monitoring of the turbulence can be performed with the copropagating local oscillator beam \cite{Elser,Heim,Peuntinger,Semenov2012}.
	As long as the classical communication time does not exceed the coherence time of the atmosphere such an approach is feasible.
	The joint PDT of our adaptive scheme $\mathcal P'$ can be obtained straightforwardly from the initial distribution $\mathcal P$ [cf. Eq.~\eqref{MomentForm}] by mapping the random variables $T_a,T_b\mapsto\min\{T_a,T_b\}$~\cite{OrderedStatistics,OrderedStatisticsBook},
	\begin{align}\label{eq:adaptivePDTC}
		\mathcal P'(T_a,T_b){=}\delta(T_a{-}T_b)\!\!\left[
			\int\limits_{T_a}^1\!\!dT'_a\mathcal P(T'_a,T_b)
			{+}\!\!
			\int\limits_{T_b}^1\!\!dT'_b\mathcal P(T_a,T'_b)
		\right]\!\!.
	\end{align}
	Thus, applying the proposed steps of this protocol, the fading channels become perfectly correlated, as $T_a=T_b$ results in $\Gamma=\Delta\Gamma=1$.
	
	At first glance it might seem counterintuitive that an additional attenuation improves the entanglement transfer.
	However, we will see that this is the case due to the resulting correlation.
	In the absence of coherent shifting, $\vec\nu=0$, we see that the Simon test for perfectly correlated channels in Eq.~\eqref{PerfectCorrWitness}, reduces to the exact form of deterministic attenuations.
	Because such a deterministic loss always preserves Gaussian entanglement \cite{Filippov,ConsatntLoss1}, this also holds true for any Gaussian entangled state with $\langle \hat{a}\rangle=0$ and $\langle \hat{b}\rangle=0$ propagating in the atmosphere and by applying our adaptive protocol.
	This is an important finding, as it shows that there is no trade-off due to artificial attenuation by using the adaptive scheme, as long as $\langle T_a^2\rangle,\langle T_b^2\rangle\neq0$.
	The additional attenuation, due to the integration in Eq.~\eqref{eq:adaptivePDTC}, is worthwhile, as the introduced correlation between the modes ($\Gamma=\Delta\Gamma=1$) assures the survival of Gaussian entanglement.

	The influence of the turbulence occurs when we consider coherent displacement which results in a non-zero second term $\vec\nu^\dagger \mathcal{S}\vec\nu$ [see Eq.~\eqref{PerfectCorrWitness}].
	Notably, this term is not necessarily positive (see previous discussion of uncorrelated channels), and is a genuine effect of turbulent loss channels which differs from deterministic loss scenarios.
	Hence, an optimal choice of coherent shifting can ensure the entanglement transfer in the atmosphere, especially for CV-QKD applications, as we discussed before.

\begin{figure}[h!]
	\includegraphics[clip=,width=1\linewidth]{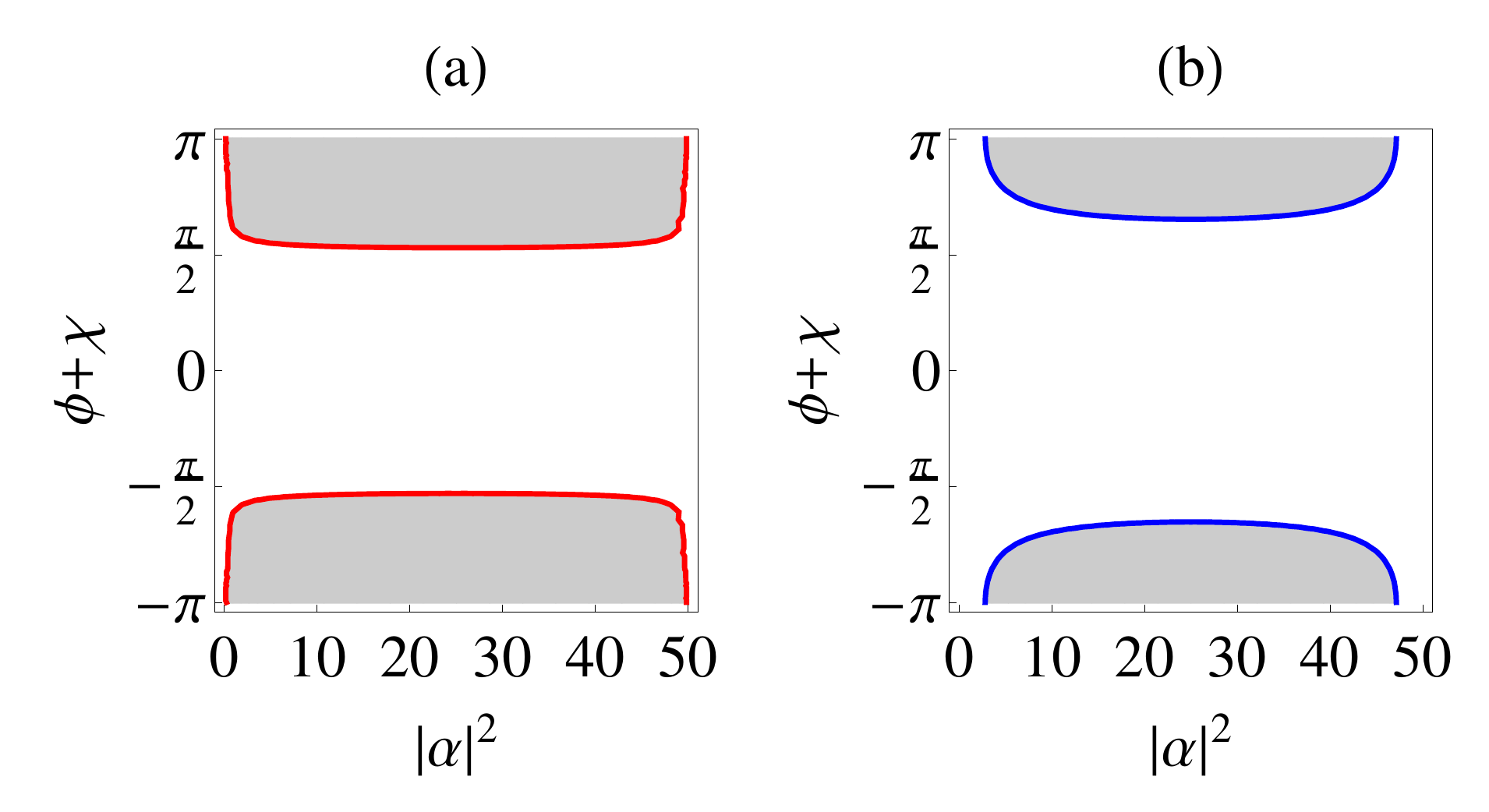}
	\caption{(Color online)
		The contours illustrate the bounds of the regions of Gaussian entanglement of a displaced TMSV state.
		Entanglement is preserved in the gray shaded areas.
		Two cases of ideal correlations $T_a{=}T_b$ $(\Delta\Gamma{=}\Gamma{=}1)$ are considered, with the same two sets of experimental turbulence parameters as in Fig.~\ref{fig:tmsv}: (a) $\langle T_a^2\rangle{=}0.163$ and $\langle T_a\rangle{=}0.398$ as well as (b) $\langle T_a^2\rangle{=}0.001$ and $\langle T_a\rangle{=}0.027$.
		The dependence on the square amplitude $|\alpha|^2{=}|\langle\hat a\rangle|^2$ and the sum of the phases of displacement $\phi{+}\chi$ is shown.
		The squared displacement amplitude of the second mode is $|\langle\hat b\rangle|^2{=}50{-}|\langle\hat a\rangle|^2$.
		The squeezing parameter is $\xi{=}0.5$. 
	}\label{fig:CoShift}
\end{figure}
	
	In Fig.~\ref{fig:CoShift} the entanglement test for a displaced TMSV state in a correlated fading channel ($T_a=T_b$) is shown for two different atmospheric characteristics.
	The state has a fixed, joint displacement amplitude $2|\vec\nu|^2=|\langle\hat a\rangle|^2+|\langle\hat b\rangle|^2=50$.
	The dependency on $|\alpha|^2=|\langle \hat a \rangle|^2$ and the sum of the phases of the coherent amplitudes $\phi+\chi$, with $\langle \hat a \rangle=|\langle \hat a \rangle|e^{i\phi}$ and $\langle \hat b \rangle=|\langle \hat b \rangle|e^{i\chi}$, is depicted.
	If Gaussian entanglement persists, it strictly depends on the choice of phase of the coherent shifting.
	In particular $\phi+\chi=0$ and $\phi+\chi=\pm\pi$ lead to the worst and the optimal case, respectively.
	Additionally, the sensitivity to the channel characteristics $\langle T_a^2\rangle$ and $\langle T_a\rangle$ can be seen by comparing the two cases.
	However, there also exists the entanglement-persisting region, which does not depend on the channel characteristics~\cite{Supplementary}.

\section{Summary and conclusions}\label{ch:summary}

	We have introduced input-output relations for the entanglement of bipartite Gaussian states propagated through the turbulent atmosphere.
	In particular, our rigorous studies demonstrate that the Gaussian entanglement preservation strongly depends on the initial coherent amplitudes, which is not the case for standard Gaussian channels.
	Moreover, we show that optimal and finite squeezing levels exist that are preferable for fading quantum communication links.
	Our findings open up new perspectives for optimal CV-QKD protocols in free-space links which employ entangled Gaussian states and encode information via coherent displacements.

	For uncorrelated fading channels, one can choose the displacement to increase the range of distributed Gaussian entanglement by steering the input state with passive optical elements.
	In addition, we proposed an adaptive technique which, by monitoring and controlling the channel transmittance, can correlate the atmospheric channel characteristics.
	In this manner, Gaussian entanglement is always preserved.
	Therefore, this approach renders it possible to distribute Gaussian entanglement between arbitrary points on Earth and orbiting satellites via atmospheric links.
	We believe that our rigorous studies and proposed methods will find a number of applications for atmospheric continuous-variable entanglement transfer, global quantum-communication networks, and in-lab experiments to improve observable properties of Gaussian entanglement in turbulent media.
	An extension of our treatment to non-Gaussian entanglement certifiers might further improve the entanglement detection and the related applications in free-space channels.

\section*{Acknowledgements}
	The authors acknowledge financial support by Deutsche Forschungsgemeinschaft through SFB 652 and VO 501/22-1.
	We are grateful to E. Agudelo for valuable comments and E. Shchukin and D. Yu. Vasylyev for enlightening discussions.

\begin{widetext}
\section*{Supplemental Material -- Gaussian entanglement in the turbulent atmosphere}

\subsection*{Determinant expansion}\label{App:Relations}
	For the following treatment of bipartite covariance matrices, let us give some well-known relations.
	Those are useful for a compact formulation of our theory.
	Firstly, relations for sums of $2\times2$ matrices and $4\times 4$ block matrices are given.
	Say $A,B,C,D\in\mathbb C^{2\times2}$ and $J=\left(\begin{smallmatrix}0&1\\-1&0\end{smallmatrix}\right)=-J^T$.
	It holds
	\begin{align}
		\det(A+B)=&\det A+\det B-{\rm tr}(AJB^TJ),\label{eq:Rel1}
		\\\nonumber \det\begin{pmatrix}A&D\\C&B\end{pmatrix}=&\det A\det B+\det C\det D
		\\
		&-{\rm tr}(AJC^TJBJD^TJ),\label{eq:Rel2}
	\end{align}
	which can be directly verified when expanding the matrices in components and comparing both sides.

	Secondly, in a two-dimensional system, the orthogonal vector $\vec\xi_\perp$ to $\vec\xi$ is given by
	\begin{align}
		\vec\xi_\perp=J\vec\xi^{\,\ast}
		\quad\text{or}\quad
		\vec\xi_\perp^{\,\dagger}=-\vec\xi^{\,T}J,
	\end{align}
	with identical lengths: $\vec\xi^{\,\dagger} \vec\xi={\vec\xi_\perp}^{\,\dagger} \vec\xi_\perp$.
	A related, useful relation for $2\times2$ matrices is $JA^TJ=-(\det A)A^{-1}$.

	From the relations~\eqref{eq:Rel1} and~\eqref{eq:Rel2}, let us deduce a rule for an Hermitian block matrix, i.e., $A=A^\dagger$ and $B=B^\dagger$, as well as $D=C^\dagger$.
	Moreover, we assume decompositions $A\mapsto A+\vec\alpha\vec\alpha^\dagger$, $B\mapsto B+\vec\beta\vec\beta^\dagger$, and $C\mapsto gC+h\vec\beta\vec\alpha^\dagger$, for $g,h\in\mathbb C$ and $\vec\alpha,\vec\beta\in\mathbb C^2$.
	Applying the previous formulas, we find
	\begin{align}\label{eq:DetFullExpansion}
	\begin{aligned}
		&\det\begin{pmatrix}
			A+\vec\alpha\vec\alpha^\dagger & g^\ast C^\dagger+h^\ast\vec\alpha\vec\beta^\dagger \\
			gC+h\vec\beta\vec\alpha^\dagger & B+\vec\beta\vec\beta^\dagger
		\end{pmatrix}\\
		=&|g|^2\det\begin{pmatrix}A&C^\dagger\\C&B\end{pmatrix}
		+(1-|g|^2)\det\begin{pmatrix}\det A&g^\ast\det C^\dagger\\g\det C&\det B\end{pmatrix}
		+(1-|h|^2)\det\begin{pmatrix}
			\vec\alpha^\dagger_\perp A\vec\alpha_\perp & g^\ast\vec\alpha^\dagger_\perp C^\dagger\vec\beta_\perp \\
			g\vec\beta^\dagger_\perp C\vec\alpha_\perp & \vec\beta^\dagger_\perp B\vec\beta_\perp
		\end{pmatrix}
		\\&+\begin{pmatrix}\vec\beta_\perp\\-\vec\alpha_\perp\end{pmatrix}^\dagger\begin{pmatrix}
			\det A(B-|g|^2CA^{-1}C^\dagger) & -g^\ast h\det C^\dagger(|g|^2C-BC^{-1 \dagger}A) \\
			-gh^\ast\det C(|g|^2C^\dagger-AC^{-1}B) & \det B(A-|g|^2C^\dagger B^{-1}C)
		\end{pmatrix}\begin{pmatrix}\vec\beta_\perp\\-\vec\alpha_\perp\end{pmatrix}.
	\end{aligned}
	\end{align}

	Let us further assume that the Hermitian $4\times4$ matrix is non-negative, i.e., $M=\left(\begin{smallmatrix}A&C^\dagger \\C&B\end{smallmatrix}\right)\geq0$.
	For any $|g|\leq 1$, one can observe that
	\begin{align}
		0\leq\begin{pmatrix} A & g^\ast C^\dagger \\ g C & B\end{pmatrix}
		=\begin{pmatrix}g^\ast&0\\0&1\end{pmatrix}^\dagger
		\begin{pmatrix} A & C^\dagger \\ C & B\end{pmatrix}
		\begin{pmatrix}g^\ast&0\\0&1\end{pmatrix}
		+(1-|g|^2)\begin{pmatrix} A & 0 \\ 0 & 0\end{pmatrix},
	\end{align}
	where both terms on the right-hand-side are non-negative.
	Moreover, the matrix $\left(\begin{smallmatrix}\det A&\det C^\dagger \\\det C&\det B\end{smallmatrix}\right)$ is non-negative, because of $\det A\geq0$, $\det B\geq0$, and
	\begin{align}
		0\leq \det M=\det A\det B+|\det C|^2-{\rm tr}\big(
			\underbrace{A}_{\geq0}
			\underbrace{\left[
				J C^T J B (J C^T J)^\dagger
			\right]}_{\geq0}
		\big)\leq\det A\det B+|\det C|^2
		=\det\begin{pmatrix}\det A &\det C^\dagger\\\det C&\det B\end{pmatrix},
	\end{align}
	applying Eq.~\eqref{eq:Rel1}.
	Finally, the matrix $\left(\begin{smallmatrix}\vec\alpha^\dagger A\vec\alpha&\vec\alpha^\dagger C^\dagger\vec\beta \\\vec\beta^\dagger C\vec\alpha&\vec\beta^\dagger B\vec\beta\end{smallmatrix}\right)$ is non-negative, since for all vectors $\left(\begin{smallmatrix}x\\y\end{smallmatrix}\right)\in\mathbb C^2$ holds
	\begin{align}
		\begin{pmatrix}x\\y\end{pmatrix}^\dagger
		\begin{pmatrix}
			\vec\alpha^\dagger A\vec\alpha & \vec\alpha^\dagger C^\dagger\vec\beta \\
			\vec\beta^\dagger C\vec\alpha & \vec\beta^\dagger B\vec\beta
		\end{pmatrix}
		\begin{pmatrix}x\\y\end{pmatrix}
		=\begin{pmatrix}x\vec\alpha\\y\vec\beta\end{pmatrix}^\dagger
		M
		\begin{pmatrix}x\vec\alpha\\y\vec\beta\end{pmatrix}
		\geq0.
	\end{align}

\subsection*{Partially transposed matrices}\label{App:Covar}
	The initial second-order matrix in bosonic creation and annihilation operators reads
	\begin{align}
		V=V_{1,1,1}=&\begin{pmatrix}
			\langle \Delta\hat a^\dagger\Delta \hat a\rangle & \langle \Delta\hat a^{\dagger2}\rangle & \langle \Delta\hat a^\dagger\Delta \hat b\rangle & \langle \Delta\hat a^\dagger\Delta \hat b^\dagger\rangle \\
			\langle \Delta\hat a^2\rangle & \langle \Delta\hat a\Delta\hat a^\dagger\rangle & \langle \Delta\hat a\Delta \hat b\rangle & \langle \Delta\hat a\Delta \hat b^\dagger\rangle \\
			\langle \Delta\hat a\Delta\hat b^\dagger\rangle & \langle \Delta\hat a^\dagger\Delta\hat b^\dagger\rangle & \langle \Delta\hat b^\dagger\Delta \hat b\rangle & \langle \Delta \hat b^{\dagger2}\rangle \\
			\langle \Delta\hat a\Delta\hat b\rangle & \langle \Delta\hat a^\dagger\Delta\hat b\rangle & \langle \Delta\hat b^2\rangle & \langle \Delta \hat b\Delta \hat b^\dagger\rangle
		\end{pmatrix}
		=\begin{pmatrix}
			A & C^\dagger \\ C & B
		\end{pmatrix},
	\end{align}
	the latter in a $2\times 2$ block-matrix form and with $\Delta\hat x=\hat x-\langle\hat x\rangle$ as well as $\langle \Delta\hat a\Delta \hat a^\dagger\rangle=\langle \Delta\hat a^\dagger\Delta \hat a\rangle+1$ and $\langle \Delta\hat b\Delta \hat b^\dagger\rangle=\langle \Delta\hat b^\dagger\Delta \hat b\rangle+1$.
	The block $A$($B$) describes the covariance of the first(second) subsystem, and $C$ includes the correlations between the subsystems.
	The index of $V_{1,1,1}$ will be explained shortly below.

	The partially transposed matrix may be decomposed in the forms
	\begin{align}
		V_{1,1,1}^{\rm PT}=&\begin{pmatrix}
			\langle \Delta\hat a^\dagger\Delta \hat a\rangle & \langle \Delta\hat a^{\dagger2}\rangle & \langle \Delta\hat a^\dagger\Delta \hat b^\dagger\rangle & \langle \Delta\hat a^\dagger\Delta \hat b\rangle \\
			\langle \Delta\hat a^2\rangle & \langle \Delta\hat a\Delta\hat a^\dagger\rangle & \langle \Delta\hat a\Delta \hat b^\dagger\rangle & \langle \Delta\hat a\Delta \hat b\rangle \\
			\langle \Delta\hat a\Delta\hat b\rangle & \langle \Delta\hat a^\dagger\Delta\hat b\rangle & \langle \Delta\hat b^\dagger\Delta \hat b\rangle & \langle \Delta \hat b^2\rangle \\
			\langle \Delta\hat a\Delta\hat b^\dagger\rangle & \langle \Delta\hat a^\dagger\Delta\hat b^\dagger\rangle & \langle \Delta\hat b^{\dagger2}\rangle & \langle \Delta \hat b\Delta \hat b^\dagger\rangle
		\end{pmatrix}
		=\begin{pmatrix}1&0\\0&X\end{pmatrix}
		\begin{pmatrix} A & C^\dagger \\ C & B \end{pmatrix}
		\begin{pmatrix}1&0\\0&X\end{pmatrix}
		-\begin{pmatrix}0&0\\0&Z\end{pmatrix}
		=\begin{pmatrix}A&C^\dagger X\\XC&B^T\end{pmatrix},
		\label{Vpt}
	\end{align}
	where the transposition is performed in subsystem $b$, and with $X=\left(\begin{smallmatrix}0&1\\1&0\end{smallmatrix}\right)$ and $Z=\left(\begin{smallmatrix}1&0\\0&-1\end{smallmatrix}\right)$.
	If we diminish the correlations in terms, $C\mapsto\Gamma C$ for $|\Gamma|<1$, we write
	\begin{align}
		V_{1,1,\Gamma}=\begin{pmatrix}
			A & \Gamma^\ast C^\dagger \\ \Gamma C & B
		\end{pmatrix}.
	\end{align}
	Moreover, we assume a constant loss,
	$
		\langle\hat a^{\dagger m}\hat a^n\hat b^{\dagger p}\hat b^q\rangle
		\stackrel{\rm loss}{\longmapsto}
		\sqrt{\eta_a}^{m+n}\sqrt{\eta_b}^{p+q}
		\langle	\hat a^{\dagger m}\hat a^n\hat b^{\dagger p}\hat b^q\rangle
	$.
	This results in the attenuated matrix
	\begin{align}\label{eq:constLossCov}
		V_{\eta_a,\eta_b,1}=\begin{pmatrix}\sqrt{\eta_a}&0\\0&\sqrt{\eta_b}\end{pmatrix}
		\begin{pmatrix} A & C^\dagger \\ C & B \end{pmatrix}
		\begin{pmatrix}\sqrt{\eta_a}&0\\0&\sqrt{\eta_b}\end{pmatrix}
		+\begin{pmatrix}(1-\eta_a)\frac{1-Z}{2} &0\\0&(1-\eta_b)\frac{1-Z}{2} \end{pmatrix}
		=\begin{pmatrix} \tilde A & \tilde C^\dagger \\ \tilde C &\tilde B 
		\end{pmatrix}^{\rm PT}.
	\end{align}
	Where we used the standard method for describing attenuations in quantum optics which is formulated in terms of beam splitters with the transmission coefficients $T_a=\sqrt{\eta_a}$ and $T_b=\sqrt{\eta_b}$.

	As it has been shown, the atmosphere can be modeled by treating $T_a$ and $T_b$ as random variables.
	This yields
	\begin{align}
		\langle \hat a^{\dagger m}\hat a^n\hat b^{\dagger p}\hat b^q\rangle
		\stackrel{\rm atm.}{\longmapsto}
		\langle T_a^{m+n}T_b^{p+q}\rangle\langle \hat a^{\dagger m}\hat a^n\hat b^{\dagger p}\hat b^q\rangle,
	\end{align}
	where the first expectation value, $\langle T_a^{m+n}T_b^{p+q}\rangle$, is given in terms of the joint probability distribution of the transmission coefficients of the atmosphere, $\mathcal P(T_a,T_b)$.
	Inserting this relation yields the matrices after the propagation in an atmospheric 
	channel as
	\begin{align}\nonumber
		V\mapsto V_{\rm atm.}=&\begin{pmatrix}
			\langle T_a^2\rangle A+(1-\langle T_a^2\rangle)\frac{1-Z}{2} + \langle \Delta T_a^2\rangle\vec\mu_a\vec\mu_a^\dagger &
			\langle T_aT_b\rangle C^\dagger + \langle \Delta T_a\Delta T_b\rangle\vec\mu_a\vec\mu_b^\dagger \\
			\langle T_aT_b\rangle C + \langle \Delta T_a\Delta T_b\rangle\vec\mu_b\vec\mu_a^\dagger &
			\langle T_b^2\rangle B+(1-\langle T_b^2\rangle)\frac{1-Z}{2} + \langle \Delta T_b^2\rangle\vec\mu_b\vec\mu_b^\dagger
		\end{pmatrix}
		\\=&
		V_{\langle T_a^2\rangle,\langle T_{b}^2\rangle,\Gamma}
		\label{eq:CovExpandLine1}
		+\begin{pmatrix}
			\vec\mu_a\vec\mu_a^\dagger &
			\Delta\Gamma \vec\mu_a\vec\mu_b^\dagger \\
			\Delta\Gamma \vec\mu_b\vec\mu_a^\dagger &
			\vec\mu_b\vec\mu_b^\dagger
		\end{pmatrix}
	\end{align}
	with $\vec\mu_a=\sqrt{\langle \Delta T_a^2\rangle}\left(\begin{smallmatrix}\langle \hat a^\dagger\rangle \\ \langle\hat a\rangle \end{smallmatrix}\right)$,
	$\vec\mu_b=\sqrt{\langle \Delta T_b^2\rangle}\left(\begin{smallmatrix}\langle \hat b^\dagger\rangle \\ \langle\hat b\rangle \end{smallmatrix}\right)$,
	and the correlation coefficients
	\begin{align}
		\Gamma=\frac{\langle T_a T_b\rangle}{\sqrt{\langle T_a^2\rangle\langle T_b^2\rangle}}
		\quad\text{and}\quad
		\Delta\Gamma=\frac{\langle \Delta T_a\Delta T_b\rangle}{\sqrt{\langle \Delta T_a^2\rangle\langle \Delta T_b^2\rangle}}.
	\end{align}

	In the first term in line~\eqref{eq:CovExpandLine1}, we have the scenario of constant losses ($\eta_x=\langle T_x^2\rangle$ for $x=a,b$) if $\Gamma=1$, otherwise, $\Gamma<1$, we have a constant loss including decreased correlations, $C\mapsto\Gamma C$.
	In second term in line~\eqref{eq:CovExpandLine1}, we have the major contribution of 
	the atmosphere in terms of fluctuations of the transmission coefficients, $\Delta 
	T_a,\Delta T_b\neq0$, and the dependence on the local displacements $\vec\mu_a$ 
	and $\vec\mu_b$.
	Note that, in the case of uncorrelated channels, $\langle T_a^mT_b^n\rangle=\langle T_a^m\rangle\langle T_b^n\rangle$, we have $\Delta\Gamma=0$ and $\Gamma=\frac{\langle T_a\rangle}{\sqrt{\langle T_a^2\rangle}}\frac{\langle T_b\rangle}{\sqrt{\langle T_b^2\rangle}}$, and for perfectly correlated links, $\langle T_a^mT_b^n\rangle=\langle T_a^{m+n}\rangle=\langle T_b^{m+n}\rangle$, we get the maximal values $\Gamma=\Delta \Gamma=1$.
	Finally, the partially transposed matrix after a propagation in an atmospheric channel 
	takes the form
	\begin{align}
		V_{\rm atm.}^{\rm PT}=V_{\langle T_a^2\rangle,\langle T_{b}^2\rangle,\Gamma }^{\rm PT}
		\label{eq:CovPTExpandLine1}
		+\begin{pmatrix}
			\vec\mu_a\vec\mu_a^\dagger &
			\Delta\Gamma \vec\mu_a\vec\mu_b^T \\
			\Delta\Gamma \vec\mu_b^\ast\vec\mu_a^\dagger &
			\vec\mu_b^\ast\vec\mu_b^T
		\end{pmatrix}.
	\end{align}
	Note that the second summand is a positive semi-definite matrix. 

\subsection*{Simon criterion}\label{App:Simon}

	The Simon entanglement test is then easily obtained by calculating the determinant $\det V_{\rm atm.}^{\rm PT}$ using Eq.~\eqref{eq:DetFullExpansion}.
	Bipartite Gaussian entanglement is revealed if and only if $\det V_{\rm atm.}^{\rm PT}<0$.
	For the case of no coherent displacement, $\langle\hat a\rangle=\langle\hat b\rangle=0$, we find
	\begin{align}\label{eq:AtmMinor0}
		\det V_{\rm atm.}^{\rm PT}=\det V_{\langle T_a^2\rangle,\langle T_{b}^2\rangle,\Gamma}^{\rm PT}
		=\Gamma^2\det V_{\langle T_a^2\rangle,\langle T_{b}^2\rangle,1}^{\rm PT}
		+(1-\Gamma^2)\underbrace{\det\begin{pmatrix}
			\det \tilde A & \Gamma\det \tilde C^\dagger\\\Gamma \det \tilde C&\det \tilde B
		\end{pmatrix}}_{=\mathcal N},
	\end{align}
	which separates the well characterized case of constant loss $\det V_{\langle T_a^2\rangle,\langle T_{b}^2\rangle,1}$ from the part that diminishes the correlations, $C\mapsto \Gamma C$.
	In the most general case, including coherent shifting in Eq.~\eqref{eq:CovPTExpandLine1} and using Eq.~\eqref{eq:DetFullExpansion}, we obtain the entanglement condition
	\begin{align}\label{eq:AtmMinor}
	\begin{aligned}
		0>\mathcal W_{\rm atm.}=&\det V_{\rm atm.}^{\rm PT}\\
		=&\Gamma^2\det V_{\langle T_a^2\rangle,\langle T_{b}^2\rangle,1}^{\rm PT}
		+(1-\Gamma^2)\det\begin{pmatrix}
			\det \tilde A & \Gamma\det\tilde C^\dagger\\\Gamma\det\tilde C&\det\tilde B
		\end{pmatrix}
		+(1-\Delta\Gamma^2)\underbrace{\det\begin{pmatrix}
			\vec{\mu}^\dagger_{a\perp}\tilde A\vec\mu_{a\perp} & \Gamma\vec\mu^\dagger_{a\perp}\tilde C^\dagger\vec\mu_{b\perp}^\ast \\
			\Gamma\vec\mu^{\rm T}_{b\perp}\tilde C\vec\mu_{a\perp} & \vec\mu^{\rm T}_{b\perp}\tilde B\vec\mu_{b\perp}^\ast
		\end{pmatrix}}_{=\mathcal F}
		\\&+\begin{pmatrix}\vec\mu_{b\perp}^\ast\\-\vec\mu_{a\perp}\end{pmatrix}^\dagger\begin{pmatrix}
			\det\tilde A(\tilde B-\Gamma^2\tilde C\tilde A^{-1}\tilde C^\dagger) & -\Gamma \Delta\Gamma\det\tilde C^\dagger(\Gamma^2\tilde C-\tilde B\tilde C^{-1\dagger}\tilde A) \\
			-\Gamma\Delta\Gamma\det\tilde C(\Gamma^2\tilde C^\dagger-\tilde A\tilde C^{-1}\tilde B) & \det \tilde B(\tilde A-\Gamma^2\tilde C^\dagger \tilde B^{-1}\tilde C)
		\end{pmatrix}\begin{pmatrix}\vec\mu_{b\perp}^\ast\\-\vec\mu_{a\perp}\end{pmatrix},
	\end{aligned}
	\end{align}
	where the last summand can also be rewritten as $\vec\nu^\dagger\mathcal S\vec \nu$.

	We can rewrite $\mathcal N$ and $\mathcal F$, defined in Eqs.~\eqref{eq:AtmMinor0} and~\eqref{eq:AtmMinor}, respectively, in the equivalent forms
	\begin{align}
		\mathcal N=\det\begin{pmatrix}
			\det\tilde A & -\Gamma\det\tilde C^\dagger \\
			-\Gamma\det\tilde C & \det\tilde B
		\end{pmatrix}=
		\det\begin{pmatrix}
			\det(\tilde A) & \Gamma\det(\tilde C^\dagger X) \\
			\Gamma\det(X \tilde C) & \det(\tilde B^T)
		\end{pmatrix},
	\end{align}
	using $\det \tilde B=\det \tilde B^T$ and $(-1)\det \tilde C=\det X\det \tilde C$, and
	\begin{align}
		\mathcal F=\det\begin{pmatrix}
			\vec{\mu}^\dagger_{a\perp}\tilde A\vec\mu_{a\perp} & \Gamma\vec\mu^\dagger_{a\perp}(\tilde C^\dagger X)\vec\mu_{b\perp} \\
			\Gamma\vec\mu^\dagger_{b\perp}(X\tilde C)\vec\mu_{a\perp} & \vec\mu^\dagger_{b\perp}\tilde B^T\vec\mu_{b\perp}
		\end{pmatrix}
	\end{align}
	using that $X\left(\begin{smallmatrix}t\\t^\ast\end{smallmatrix}\right)=\left(\begin{smallmatrix}t\\t^\ast\end{smallmatrix}\right)^\ast$ and $X^2=\left(\begin{smallmatrix}1&0\\0&1\end{smallmatrix}\right)$.
	The right-hand-side formulations show that $\mathcal F$ and $\mathcal N$ are invariant under partial transposition [cf. Eq.~\eqref{Vpt}] and, therefore, non-negative.
	The latter follows from expanding $\det V_{\rm atm.}$, which is non-negative; see also the discussion on non-negativity in Sec.~\ref{App:Relations}.

\subsection*{Channel-independent entanglement condition for correlated channels}\label{SupplementaryB}
	Let us consider the $2{\times} 2$ matrix $D^{\rm PT}$, which is built from the elements of 
	the first and third rows and columns of the matrix $V_{1,1,1}^{\rm PT}$ in Eq.~\eqref{Vpt},
	\begin{align}
		D^{\rm PT}=\begin{pmatrix}
			\langle\Delta \hat{a}^\dag \Delta \hat{a} \rangle
			& \langle\Delta \hat{a}^\dag \Delta \hat{b}^\dag \rangle
			\\ \langle\Delta \hat{a} \Delta \hat{b} \rangle
			& \langle\Delta \hat{b}^\dag \Delta \hat{b} \rangle 
		\end{pmatrix}.\label{DuanMatrix}
	\end{align}
	The negative determinant of this matrix, $\det D^{\rm PT}<0$, is a sufficient condition of the 
	Gaussian entanglement.
	This condition corresponds to the Duan \textit{et al.} criterion.
	Now, we assume a perfectly correlated channel, $T=T_a=T_b$. 
	Performing the same consideration like in the case of the Simon criterion, one gets for the 
	correlated channels, 
	\begin{align}
		\det D^{\rm PT}_\mathrm{atm.}
		=\langle T^2\rangle^2\det D^{\rm PT}
		+\langle\Delta T^2\rangle \langle T^2\rangle
		\begin{pmatrix}\langle\hat a\rangle\\-\langle\hat b^\dagger\rangle\end{pmatrix}^\dagger
		\begin{pmatrix}
			\langle\Delta\hat{b}^\dag\Delta\hat{b}\rangle
			&\langle\Delta\hat{a}\Delta\hat{b}\rangle
			\\ \langle\Delta\hat{a}^\dag\Delta\hat{b}^\dag\rangle
			& \langle\Delta\hat{a}^\dag\Delta\hat{a}\rangle
		\end{pmatrix}
		\begin{pmatrix}\langle\hat a\rangle\\-\langle\hat b^\dagger\rangle\end{pmatrix}.
	\end{align}
	The negativity of the second term,
	\begin{align}
		\begin{pmatrix}\langle\hat a\rangle\\-\langle\hat b^\dagger\rangle\end{pmatrix}^\dagger
		\begin{pmatrix}
			\langle\Delta\hat{b}^\dag\Delta\hat{b}\rangle
			&\langle\Delta\hat{a}\Delta\hat{b}\rangle
			\\ \langle\Delta\hat{a}^\dag\Delta\hat{b}^\dag\rangle
			& \langle\Delta\hat{a}^\dag\Delta\hat{a}\rangle
		\end{pmatrix}
		\begin{pmatrix}\langle\hat a\rangle\\-\langle\hat b^\dagger\rangle\end{pmatrix}<0
	\end{align}
	identifies the entanglement-persisting region, which does not depend on the channel characteristics.

\end{widetext}

\begin{thebibliography}{99}
\bibitem{OrderedStatistics}
	From a joint probability density $\mathcal P$, we get the cumulative probability distribution for $Z=\min\{X,Y\}\leq z$ as ${\rm Pr}(X{\leq} z \vee Y{\leq} z)=1-\int_z dx'\int_z dy'\mathcal P(x',y')$.
	The derivative gives the desired probability density: $\mathcal P(z)=(d/dz){\rm Pr}(X{\leq} z \vee Y{\leq} z)=\int_z dy'\mathcal P(z,y'){+}\int_z dx'\mathcal P(x',z)$.
\end{thebibliography}
\end{document}